# On the appearance of compact objects at radio and optical frequencies

**G. Chapline**
Lawrence Livermore National Laboratory, Livermore, CA 94550

**Abstract**
**In the dark energy star picture a compact object is transparent at radio and optical frequencies, and acts as a defocusing lens. Thus the object itself appears as a luminous disk whose surface brightness reflects the surface brightness of the background. In the case of rotating dark energy stars the image will also contain background independent emission features. In this note we provide simple analytic formulae for the separation of these features as a function of angular momentum and viewing angle. In the case of rapid rotation these features will appear to lie within the shadow expected if the compact object were a black hole.**

Black hole solutions of the classical Einstein equations pose a number of conceptual difficulties, not the least of which is incompatibility with elementary quantum mechanics. Some time ago it was suggested that in reality the interior of a compact object is a "squeezed" version of the ordinary space-time vacuum [Chapline 1991]. This led to the suggestion [Chapline, et. al. 2001, Mazur and Mottola 2004] that event horizons do not exist, and instead compact objects have a physical surface that produces observable effects. For example, nucleons falling onto the surface of a dark energy star will disintegrate, leading to the emission of positrons and gamma rays [Barbieri and Chapline 2004]. On the other hand an elementary particle whose momentum far from the compact object is less than a keV/c will pass through the surface unimpeded – at least in the "dark energy star" picture where the surface is a quantum critical layer [Chapline et. al 2001]. Therefore in the dark energy star picture compact objects will be transparent at radio and optical wavelengths. This means that viewed against a bight background the dark energy star will not appear to be black. Another prediction, which is the focus of this note, is that in the dark energy star picture rotating compact objects can themselves be bright sources of radio and optical radiation.

The interior of a dark energy star locally resembles de Sitter's "interior" cosmological solution [Chapline et. al. 2001]. As was noted by de Sitter in his original paper [de Sitter 1917] particles in such a space are subject to a constant outward acceleration. This means that particles that enter the interior of a dark energy star follow hyperbolic-like trajectories and emerge from the interior along a path that is the mirror image of their incoming trajectory [Chapline 2003]. As a consequence to a distant observer a dark energy star is not black, but will have a surface brightness that reflects the background. In the case of an extended uniform background the surface brightness of the dark energy star will be the same as the background. In any case there will be no "black shadow", as predicted for black holes [Bardeen 1973].

Although dark energy stars are transparent for elementary particles whose momenta a long distance from the compact object is small, it is possible for even low frequency radiation to be trapped in the Godel-like core of a rotating dark energy star [Chapline and Marecki, 2008]. In Godel-like space-times particles follow trajectories that resemble those of charged particles in a magnetic field [Novello, Soares, and Tiomno 1983]. In particular, particles in a Godel-like space-time are confined inside a cylindrical boundary, but can flow unimpeded both around and parallel to the axis of rotation. In this note we will consider the emission of radio waves coming directly from a dark energy star as a result of the expectation that highly collimated jets containing both charged particles and photons will emerge from the polar regions of a rotating dark energy star along the axis of rotation. In particular, we point out in the following that the radio and optical radiation from these jets provides a characteristic signature for dark energy stars that is quite distinct from what is expected for a black hole.

Actually it is somewhat enigmatic that recent improvements in imaging the SgA* radio source near to the center of our galax using mm wavelength VLBI [Doeleman et al 2008] suggest that the compact object itself might be a source of radio emissions. In the conventional picture where such objects are "black holes" no such emissions should be possible, since sources of photons coming very close to a black hole are sucked inside the event horizon, and fade from view. It is expected that this will result in a dark shadow with a diameter approximately equal to $10GM/c^2$, where $M$ is the mass of the black hole [Bardeen 1973]. In the case of the compact object located near to the center of our galaxy with an estimated mass of $4 \times 10^6$ solar masses [Ghez, A., et. al., 2005], the expected black hole shadow would have an angular size of approximately 50 μarcsec [Doeleman et al 2008]. Actually photons can come relatively close to the event horizon of a very rapidly rotating black hole if they are incident close to the equatorial plane in the same direction as the direction of rotation [Bardeen 1973]. This could result in a background dependent bright spot at one point on the edge of the black hole shadow [Falcke. Melia, and Agol 2000]. However, as we shall now show the background independent optical or radio image of a rotating dark energy star will have a quite different appearance.

Although the interior metric of a rotating dark energy star is quite different from the Kerr metric [Chapline and Marecki, 2008], the exterior metric should resemble the Kerr metric, particularly as one moves away from the horizon. Therefore one can use the Kerr metric to understand at least qualitatively the fate of photons emitted by the jets. The equations describing photon trajectories in a Kerr space-time are well known [Carter 1968]. Near to the polar axis and using the Boyer-Lindquist form of the metric [Boyer and Lindquist 1967] these equations have the form (in units such that $G/c^2 = 1$):

$$\frac{dr}{dt} = \frac{\pm \Delta}{(r^2 + a^2)} \left(1 - \frac{\Delta}{(r^2 + a^2)^2} b^2\right)^{1/2}$$

$$\frac{d\theta}{dt} = \frac{\Delta b}{(r^2 + a^2)^2} \qquad (1)$$

$$\frac{d\phi}{dt} = \frac{2aMr}{(r^2 + a^2)^2},$$

where $\Delta \equiv r^2 - 2Mr + a^2$, $a$ is the angular momentum per unit mass of the compact object, and $b$ is a constant. The radial speed of a photon approaching the horizon as measured by a local inertial observer on the polar axis will be:

$$\left(\frac{dr}{dt}\right)^2_{local} = 1 - \frac{\Delta b^2}{(r^2 + a^2)^2} \quad (2)$$

A light ray emitted from a source located on the polar axis will be neither deflected upwards or downwards as long as there is a value for $r$ such that the right hand side of eq. (2) vanishes. The height $z^*$ on the axis of rotation where this happens can be found by setting the derivative with respect to $r$ of the second term on the r.h.s. of eq. (2) equal to zero. This yields the following equation for $z^*$:

$$(z^*)^3 - 3M(z^*)^2 + a^2 z^* + a^2 M = 0 \quad (3)$$

This equation can be solved using the method of Cardano giving:

$$z^* = M\left\{1 + 2\left(1 - \frac{a^2}{3M^2}\right)^{1/2} \cos\left[\frac{1}{3}\cos^{-1}\frac{1 - \frac{a^2}{M^2}}{\left(1 - \frac{a^2}{3M^2}\right)^{3/2}}\right]\right\}. \quad (4)$$

In the limit $a = 0$ one obtains $z^* = r^* = 3M$, while in the limit of maximum rotation $a = M$ one obtains $z^* = (1+\sqrt{2})M$. The corresponding value of $b$ where photons are emitted horizontally without upward or downward deflection can be found by setting the r.h.s. of eq. (2) = 0 when $r = z^*$. When $a = 0$ one finds the well know result for the apparent critical radius $b^* = 3\sqrt{3}M$, while in the limit of extreme rotation $a = M$ one obtains $b^* = 2(1+\sqrt{2})M$. An simple approximate formula for $b^*$ that is accurate to 20% over the entire range of rotation is:

$$b^* \approx 2(1 + \sqrt{2})\left(1 + \frac{1}{\sqrt{2}}(1 - \frac{a^2}{M^2})\right)^{1/2} M. \quad (5)$$

The constant *b* in Eq. (1) is related to the angular momentum of a photon as it crosses the axis of rotation: $L_\theta(\theta=0) = h\nu b$. When $a = 0$ conservation of angular momentum implies that *b* is equal to the impact parameter of the photon as measured by any distant observer. However when $a \neq 0$ the components of angular momentum perpendicular to the polar axis are not conserved because the Kerr metric is not spherically symmetric, and *b* is not equal to the asymptotic impact parameter. Instead it can be shown [Carter 1968] that for a photon that passes through the axis of rotation the quantity $(L_\theta/h\nu)^2 - a^2\cos^2\theta$ is a constant of motion, where $L_\theta$ is the component of photon angular momentum perpendicular to the plane formed by the polar axis and the photon trajectory, and $h\nu$ is the photon energy. The constancy of $(L_\theta/h\nu)^2 - a^2\cos^2\theta$ implies that the impact parameter seen by a distant observer for a photon emitted from a source near to the axis of rotation will be:

$$b_\infty = (b^2 - a^2 \sin^2\theta_V)^{1/2} \quad , \tag{6}$$

where $\theta_V$ is the angle between the axis of rotation and the direction to the distant observer. For a photon emitted at the critical height $z^*$, a distant observer will see the photon emerging at an apparent radius $b_\infty^* = (b^{*2} - a^2\sin^2\theta_V)^{1/2}$ as illustrated in the figure below. In the case of maximum rotation $a = M$, an observer viewing the compact object along a direction close to the equatorial plane ($\theta_V \approx \pi/2$) would see a photon emitted near to $z^*$ emerging at a radius $b_\infty^* = 4.7M$.

The diameter of jets emerging from inside a rotating dark energy star will rapidly expand as they emerge from the surface of the dark energy star due to the fanning out of the gravitational "magnetic" field. Consequently the intensity of radio and optical emissions from the jets will rapidly decrease as a function of distance from the dark energy star. Since the photons emitted at heights below $z^*$ will tend to be gravitationally trapped, the apparent jet luminosity will be greatest in the neighborhood of $z^*$. Due to gravitational focusing near to $z^*$ this luminosity will be maximal when viewed from a direction close to the equatorial plane. A similar effect has been noted for the observed luminosity of stars orbiting close to a rotating black hole [Cunningham and Bardeen 1972]. Thus when observed from a direction close to the equatorial plane the most prominent feature of the radio or optical image of a rapidly rotating compact object will be two bright spots separated in angle by the angle subtended by $9.4GM/c^2$. For the compact object at the center of our galaxy this corresponds to an angular separation of 44 µarcsec. This is smaller than the 50 µarcsec diameter expected for the black hole shadow of a $4\times10^6$ solar mass black hole. It is, of course, a celebrated fact that it is not possible for a static source of photons to be located so close to a black hole. It is theoretically possible to have objects orbit a rotating black hole at an apparent radius of $4.7M$ if they orbit in the equatorial plane in the same direction as the direction of rotation. However, the luminosity of such objects would pulsate with a frequency equal to the orbital frequency [Cunningham and Bardeen 1972]. It should be relatively easy to exclude this possibility by measuring the time dependence of the SgA* radio emissions. Finally if the compact object is viewed from a direction close to the axis of rotation one should see a spot near the center of the image due to emission from the jet flowing directly toward the observer. In addition to the central spot the jet on the other side of the compact object should produce an "Einstein ring" with an apparent diameter $4(1+\sqrt{2})M$.

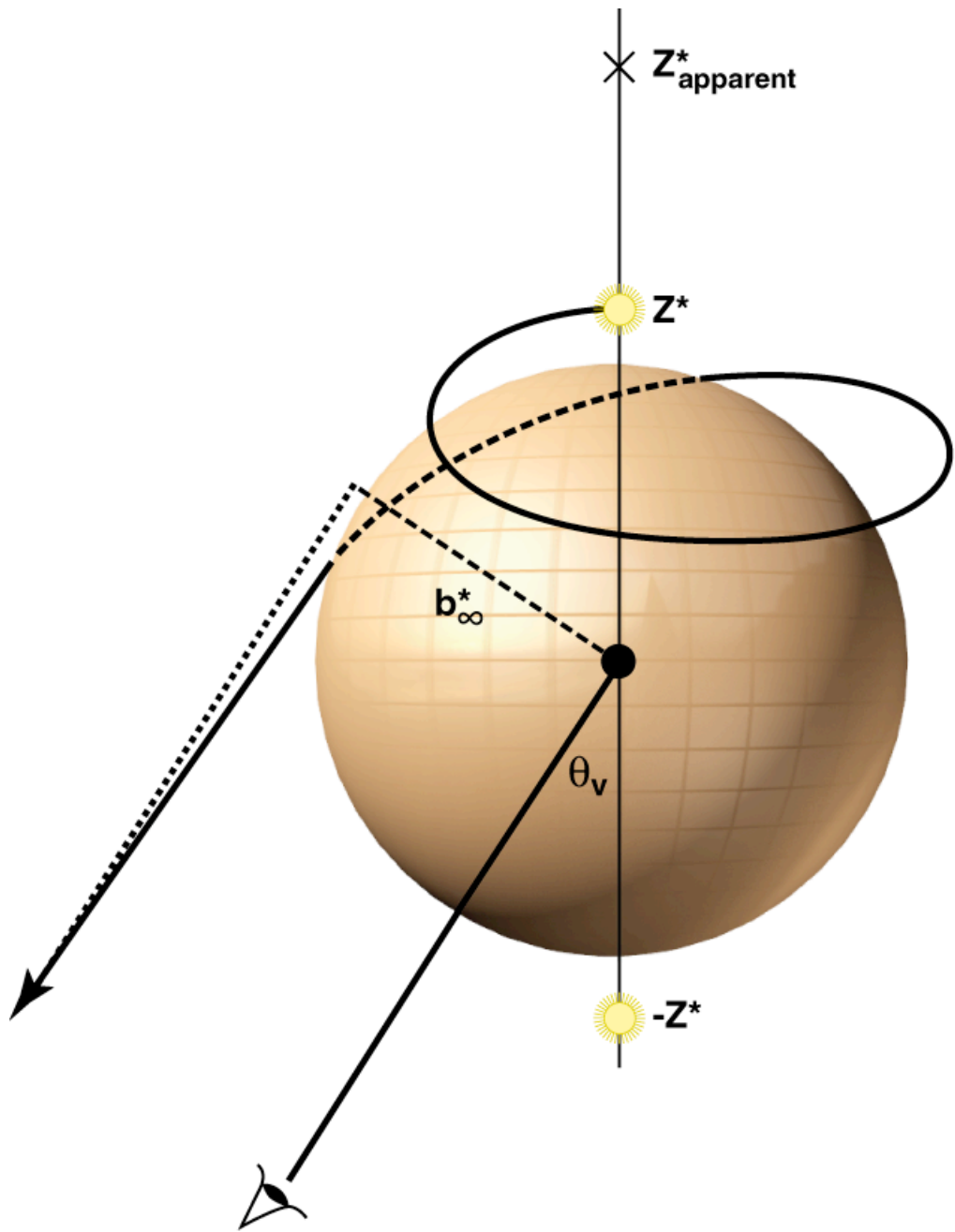

**Figure. Origin of the radio/optical emission from a rotating dark energy star**

**Acknowledgments**

The author is very grateful to Jim Barbieri, Pawel Mazur, and Piotr Marecki for discussions. This work performed under the auspices of the U.S. Department of Energy by Lawrence Livermore National Laboratory under Contract DE-AC52-07NA27344.